# The Canonical Ensemble and the Central Limit Theorem.

D. Sands and J. Dunning-Davies,
Department of Physics,
University of Hull,
Hull,
England.

email:    d.sands@hull.ac.uk
          j.dunning-davies@hull.ac.uk

**Abstract.**

Some of the more powerful results of mathematical statistics are becoming of increasing importance in statistical mechanics. Here the use of the central limit theorem in conjunction with the canonical ensemble is shown to lead to an interesting and important new insight into results associated with the canonical ensemble. This theoretical work is illustrated numerically and it is shown how this numerical work can form the basis of an undergraduate laboratory experiment which should help to implant ideas of statistical mechanics in students' minds.

# Introduction.

It is becoming more and more apparent that the development of the area of physics known as statistical mechanics independently of mathematical statistics itself was a totally unfortunate occurrence. More recently, with the work of such as Jaynes [1], this division into two apparently separate subjects has been blurred somewhat. However, the separate approaches and different terminology still tend to impose an unrealistic and artificial barrier, which can prove a real hindrance to students and new workers in the field alike. An attempt to rectify this unfortunate situation is provided by the book on statistical physics by Lavenda [2]. Even here, though, it sometimes proves difficult for newcomers to the field to appreciate the powerful techniques that actual mathematical statistics can bring to bear on problems in statistical mechanics. Nowhere is this more apparent than in the lack of knowledge of, and lack of appreciation of, the important limit theorems of mathematical statistics.- particularly the so-called 'laws of large numbers' and the 'central limit theorems'.

# The central limit theorem.

A classic example of the use of well-known results from mathematical statistics is provided by the canonical ensemble, for which the proper probability density is given by

$$f(E;\beta) = \frac{e^{-\beta E}}{Z(\beta)} \Omega(E) \quad (1)$$

where $\Omega(E) = \frac{C^m E^{m-1}}{\Gamma(m)}$, $C$ being a constant depending on the system under consideration and $2m$ being the number of degrees of freedom. Here also, $Z(\beta)$ is given by

$$Z(\beta) = \int_0^\infty e^{-\beta E} \Omega(E) dE. \quad (2)$$

It follows that the mean value of the energy is given by

$$\overline{E} = -\frac{\partial}{\partial \beta} \ln Z = \frac{1}{Z(\beta)} \int_0^\infty E e^{-\beta E}(E) dE \quad (3)$$

and

$$-\frac{\partial \overline{E}}{\partial \beta} = -\frac{\partial}{\partial \beta}\left(-\frac{\partial}{\partial \beta} \ln Z\right) = \frac{1}{Z}\frac{\partial^2 Z}{\partial \beta^2} - \frac{1}{Z^2}\left(\frac{\partial Z}{\partial \beta}\right)^2 \quad (4)$$

$$= \frac{1}{Z} \int_0^\infty E^2 e^{-\beta E} \Omega(E) dE - \left(\overline{E}\right)^2 \quad (5)$$

$$= \overline{E^2} - \left(\overline{E}\right)^2 = \overline{\left(E - \overline{E}\right)^2} = \overline{(\Delta E)^2}, \quad (6)$$

which is the variance of the distribution.

All the above is quite well-known and many of the results will be familiar. However, within the realms of the theory of statistics itself, specifically in probability theory, some of the most important results are so-called limit theorems. Among these, possibly the most useful are grouped together as either 'laws of large numbers' or 'central limit theorems'. Theorems are often termed 'laws of large numbers' if they are concerned with giving conditions under which the average of a sequence of random variables converges in some sense to the expected average. The central limit theorems, on the other hand, are concerned with finding the conditions under which the sum of a large number of random variables has a probability distribution which approximates to the normal distribution. Within statistics, the properties of the so-called 'normal' distribution have been studied in great depth. This is due in part to the fact that, in practice, many variables are distributed in accordance with this distribution or very nearly according to it. It is found that the sample means approximately follow normal distribution even when the underlying distribution is not normal. This result provides the subject matter for a central limit theorem, initially introduced by de Moivre and later improved by Laplace, although the first truly rigorous proof was presented by Lindeberg towards the beginning of the last century. This powerful theorem states that

If there are $x_1, x_2, \ldots, x_n$ identically distributed random variables, each with mean $\mu$ and finite variance $\sigma^2$, then the variable $\bar{x}$, given by

$$\bar{x} = \frac{1}{n}\sum_{i=1}^{n} x_i \qquad (7)$$

will have a distribution

$$z = \frac{\bar{x} - \mu}{\sigma/\sqrt{n}}, \qquad (8)$$

which approaches the standard normal distribution with mean 0 and variance 1 as $n$ becomes indefinitely large

The proof of this theorem may be found in most standard statistics texts, for example those by Rényi [3] and Cramer [4]. However, from a purely practical viewpoint, the importance of this theorem resides in the fact that it allows the use of results associated with the normal distribution even when the basic variable, $x$, has a distribution which differs markedly from normality. It might be noted that the more the basic distribution differs from normality, the larger $n$ must become to approximate normality for $\bar{x}$. Almost the only restriction imposed on the underlying distribution is that the variance be finite, It is encouraging to realise that the vast majority of physics problems obey this restriction quite naturally.

In the present context, it may be noted, therefore, that the distribution associated with the canonical ensemble may be approximated by

$$\frac{\exp\left\{-\frac{(E-\bar{E})^2}{2(\Delta E)^2}\right\}}{\left[2\pi\overline{[\Delta E]^2}\right]^{1/2}}. \qquad (9)$$

This somewhat surprising result, showing that the canonical distribution is effectively normal, is relatively unknown in statistical mechanics, although it is stated quite clearly in the book by Lavenda [2].

In the orthodox view normally given in undergraduate texts [5-7], the canonical distribution is expressed as

$$p_i = \frac{1}{Z} e^{-E_i/kT} \qquad (10)$$

where $p_i$ is interpreted as the probability that the system is in an energy state $E_i$. Gibbs called this, "the canonical distribution is phase" [8]. He derived it by constructing an ensemble of systems, each containing $N$ particles, but with a different distribution of velocities. Each system in the ensemble is represented by a point in $6N$-dimensional space. The Liouville theorem shows that the density of such points is constant over time, which places a constraint on the energy dependence of the phase space density. Gibbs argued leads naturally to the exponential as the canonical distribution in phase.

More modern derivations of (10) use a very different approach [5-7]. Rather than constructing an ensemble of identical systems, a small system is placed in contact with an ideal reservoir and the accessible states of the composite system are considered. There are two key assumptions. First, the total energy ($E$) is fixed but fluctuations between the reservoir ($E_R$) and the small system ($E_i$) mean that

$$E_i = E - E_R \qquad (11)$$

Second, as the reservoir is much larger than the small system the accessible states of the reservoir dominate the distribution, and the probability of finding the small system in a given energy state $E_i$ depends on the probability of finding the reservoir in a state $E_R$. It is straightforward then to show that the probability distribution takes the exponential form in (10) with $p_i$ again representing the probability of finding the system in a particular energy state.

Comparison of (9) and (10) shows that the application of the central limit theorem fundamentally changes our physical understanding of the statistics of canonical systems. Baierlein has written of (10); "Qualitatively stated, the probability of high-energy states is exponentially smaller than the probability of low-energy states". The converse must follow; the probability of low-energy states is exponentially higher then the probability of high-energy states. Therefore the most probable state is $E_i=0$, regardless of temperature. In other words, (10) implies that all the energy should reside in the reservoir, which is physically nonsensical. The canonical system has a well defined macroscopic energy given by

$$\langle E \rangle = \int_0^{T_R} c_v \, dT \qquad (12)$$

It would be expected intuitively that, if fluctuations occur, then the probability of observing both higher and lower energies would be equal. This is in fact what the normal distribution of (9) implies. The most probable state corresponds to the mean energy of the distribution and the energy can fluctuate with equal probability in either direction.

## Numerical simulation

The fact that the canonical distribution may be approximated by the normal distribution can be demonstrated using a very simple numerical simulation of a small collection of particles of an ideal gas. In fact, the assumption of ideality is not necessary. The only requirement is that the velocities assigned to the particles be drawn from a Maxwell-Boltzmann distribution, as this at least ensures that the simulation has a sound physical basis. Furthermore, the algorithm is simple enough to be implemented by any undergraduate with a knowledge of programming, as no special techniques are employed other than the generation of a uniform distribution of random numbers. Such generators are used extensively in Monte Carlo simulations and are discussed by Binder [9].

Let the gas contain $M$ particles. A sequence of $M$ velocities drawn randomly from a Maxwell-Boltzmann distribution will have a mean square speed

$$E_1 = \overline{v^2} = \frac{1}{M} \sum_{j=1}^{M} v_j^2 \tag{13}$$

Physically, the mass of the particle must be known to relate $E_1$ to the mean energy per particle but computationally it is not necessary to specify such detail and $E_1$ can therefore be set equal to the mean energy per particle. This means, of course, that it is not possible to specify the absolute temperature, but the most probable velocity in the Maxwell-Boltzmann distribution, $v_m$, can be determined and $v_m^2$ can be taken as a measure of the temperature. Another sequence of $M$ velocities drawn at random will give rise to another value of $E_1$ and for a sequence of $K$ such trials the mean energy per particle will be

$$\overline{E} = \frac{1}{K} \sum_{j=1}^{K} E_1(j) \tag{14}$$

It is an elementary property of statistics that $\overline{E}$ is equal to the mean square speed of the distribution from which the speeds are drawn, which is $3/2.v_m^2$. $\overline{E}$ is, therefore, also a measure of the temperature.

Computationally, the canonical distribution can be mapped out as the algorithm proceeds by defining an interval $\delta E$ and recording the number of times, $N(E)$, that $E_1$ lies between $E$ and $E+\delta E$. The speed distribution can be mapped out in a similar manner by defining a small interval $\delta v$. Each velocity is generated by vector addition of three separate random numbers drawn from a normal distribution (see appendix) to represent the velocities in the $x$, $y$, and $z$ directions. Then, the number of times, $N(v)$, that the randomly generated velocities

$$v = (v_x^2 + v_y^2 + v_z^2)^{\frac{1}{2}} \tag{15}$$

lie between $v$ and $v+\delta v$ defines the distribution.

In a real thermodynamic system, $M$ will typically be of the order of $10^{19}$, but this cannot be simulated realistically. $M$ must be large enough for the statistical concepts to be valid and values ranging from 100 to 4000 have been used with $K$ typically $10^5$. Four Maxwell-Boltzmann distributions have been generated with specified $v_m$=379, 456, 522, and 580 ms$^{-1}$.

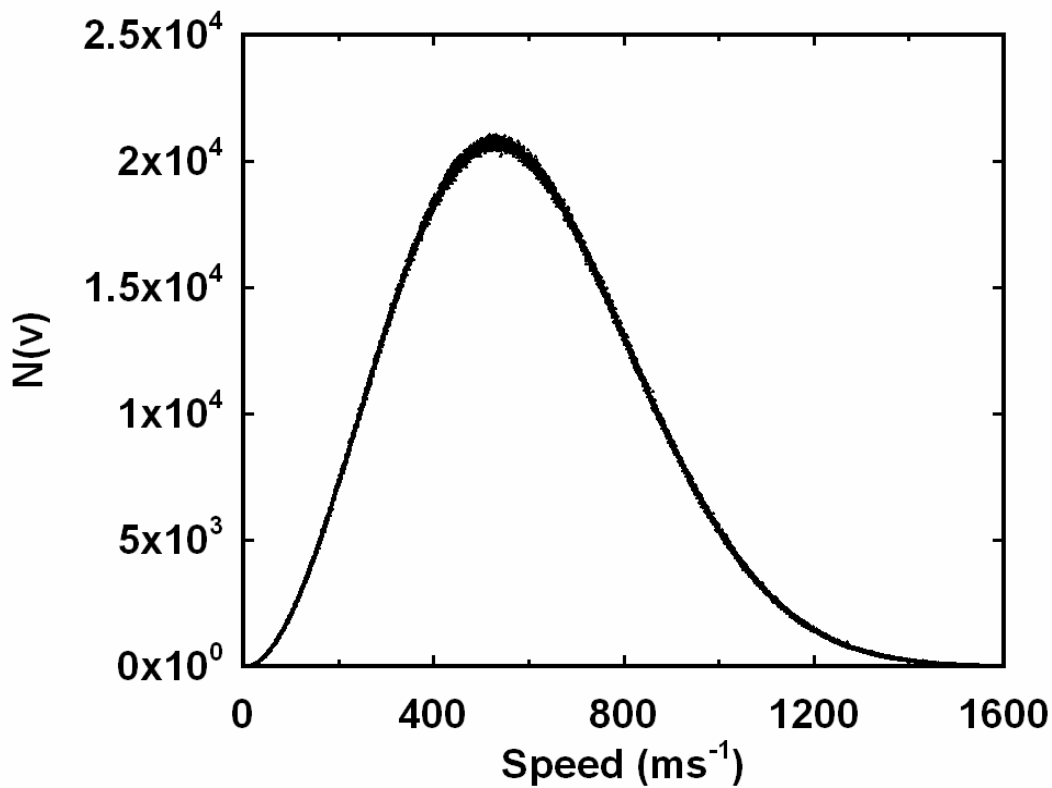

Figure 1
The Maxwell-Boltzmann speed distribution with a most probable speed of 525 ms$^{-1}$ generated from $10^8$ random numbers.

Figure 1 shows the third of these distributions generated from a total of $10^8$ velocities, which is about the limit of a simple random number generator on a 32-bit computer. The computer used in these simulations is based on a 64-bit Sun Sparc station and therefore has a much longer repeat period. The "experimental" most probable velocity derived by fitting the Maxwell-Boltzmann function to the distribution is 525 ms$^{-1}$, a difference of only 0.5%.

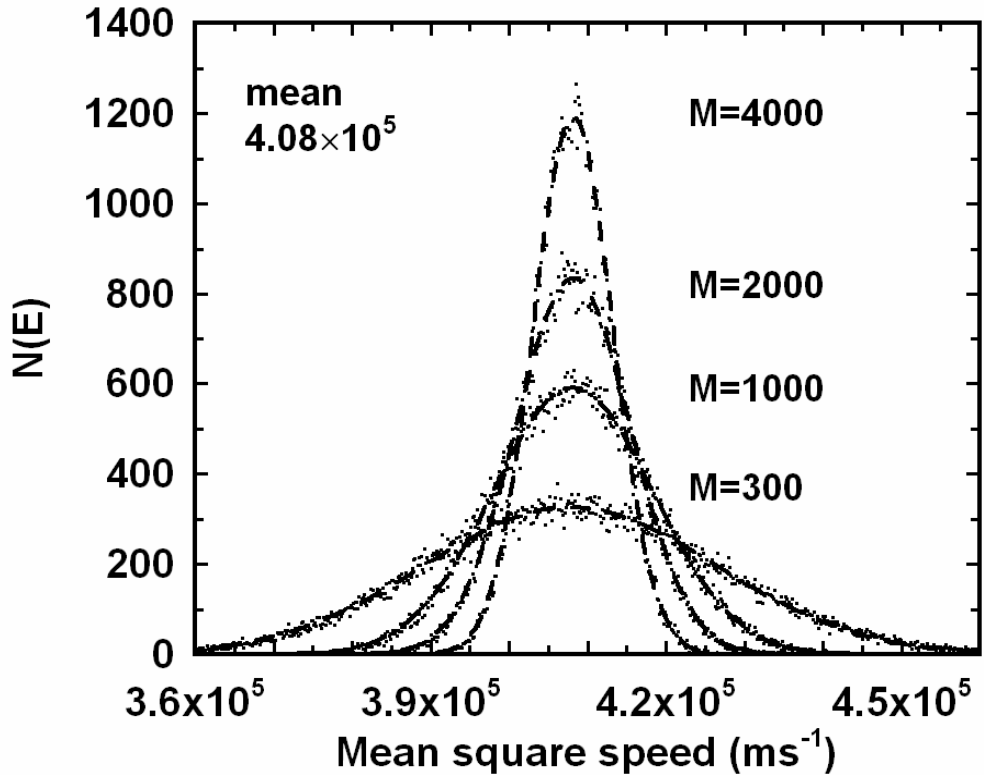

Figure 2.
The distribution of mean square speed per particle (points) together with best-fit normal distibutions (dashed lines) for the Maxwell-Boltzmann distribution in figure 1.

Figure 2 shows the distribution of mean energies per particle, $E_1$, for four values of $M$. The points represent the numerically generated distribution and the dashed lines represent the best fit from a normal distribution. The means of all the different $M$-particle canonical distributions all coincide at $\overline{E} = 4.08 \times 10^5$ (ms$^{-1}$)$^2$, which is within 1.5% of the experimentally derived $3/2 v_m^2$. A similar accuracy is observed for all four Maxwell-Boltzmann distributions generated in this simulation. Figure 2 therefore demonstrates very clearly the central limit theorem in practice: the values $E_1$, which are averages of the non-normally distributed quantities $v^2$, are themselves distributed normally.

Aside from demonstrating the central limit theorem, this simulation also provides an insight into fluctuations and the ergodic principle. The Maxwell-Boltzmann distribution is often treated as a particle distribution in statistical mechanics, which is to say that the probability, $p(v)dv$, of finding a speed lying between $v$ and $v+dv$ is accurately described by the ratio $N(v).dv/N$. This can only be the case for large systems. It is apparent from figure 2 that the width of the normal distribution decreases as $M$ increases and figure 3 shows in fact that the standard deviation varies as the inverse of $\sqrt{M}$.

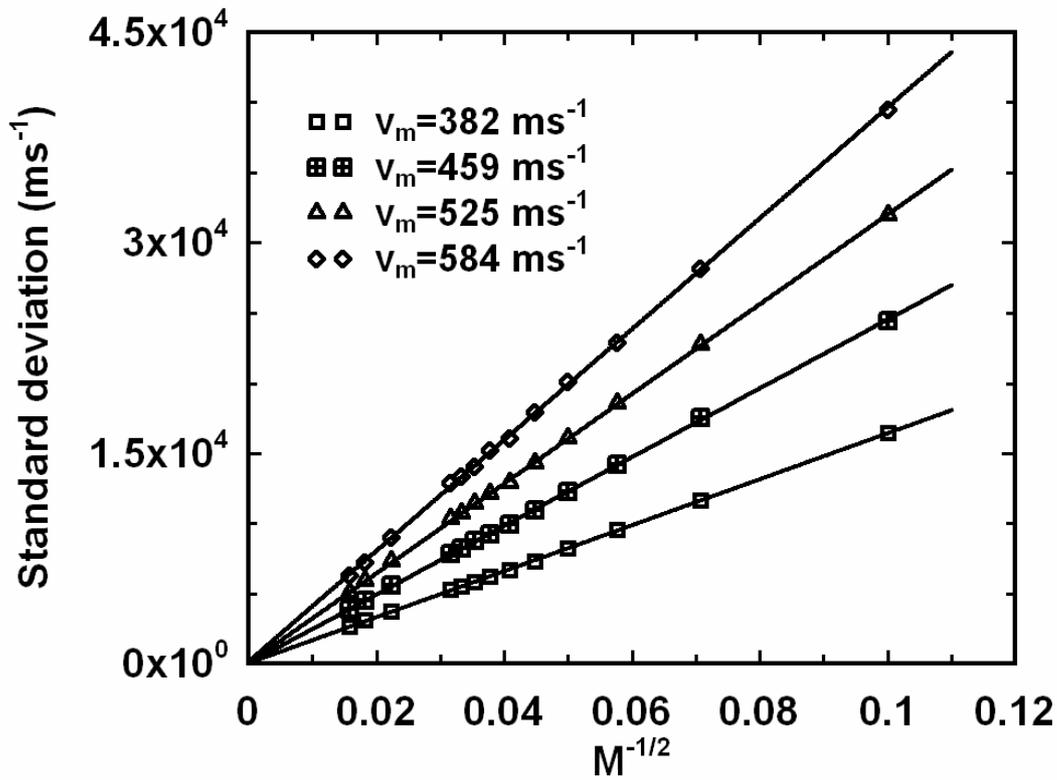

Figure 3.
The standard deviation of the normal distributions generated at the four most probable speeds shown against the inverse root of the particle number (M).

For a system containing a realistic $10^{19}$ atoms as opposed to the maximum number of 4000 simulated here, the standard deviation will be of the order of $10^{-3}$ $(ms^{-1})^2$. There is thus a negligible probability of finding the system in a state with a mean square speed different from the mean square speed of the Maxwell-Boltzmann distribution, which means that the distribution of speeds among the particles will itself be Maxwell-Boltzmann.

The same is not true for a small system where there is a significant probability of finding the system in a state with a mean square speed different from $3/2$ $v_m^2$. The distribution of particle speeds in such a state will clearly not be Maxwell-Boltzmann, even though the speeds are drawn from such a distribution. It follows then that the Maxwell-Boltzmann distribution can only be a valid as a description of the probability of finding a given speed distribution in the sense of an average. A small system similar to those simulated here might be realised by a small number $M_A$ of particles of type $A$ mixed in amongst a much, much larger number $M_B$ of particles of type $B$. The temperature of the system will be determined by the properties of $B$, which will act as the reservoir. Collisions between particles of $A$ and $B$ will ensure a constant interchange of energy between the system and the reservoir, and dynamically the small system $A$ will move through the states of the canonical distribution shown in figure 2. The instantaneous mean square speed will not necessarily be the same as the mean square speed of the Maxwell-Boltzmann distribution but, taken over time, the average mean square speed will be so. This is the ergodic principle; the time average behaviour of a small number of particles is equivalent to the instantaneous average over a very large number particles.

The canonical distribution can, therefore, be identified physically with the occurrence of fluctuations. It is well known in statistical mechanics that for a system with well-defined temperature T the energy fluctuates with a variance

$$\langle E_M \rangle^2 = kT^2 \frac{\partial U}{\partial T}, \tag{16}$$

where $U$ is the internal energy equivalent to $\overline{E}$ in the present formulation. $\langle E_M \rangle^2$ is the variance of the *M* particle distribution:

$$\langle E_M \rangle^2 = M^2 \langle E_1 \rangle^2 \tag{17}$$

For an ideal gas,

$$\frac{\partial U}{\partial T} = \frac{3}{2} k.M \tag{18}$$

Hence,

$$\langle E_1 \rangle^2 \propto \frac{1}{M} \tag{19}$$

and the standard deviation varies as the inverse of √M, which is confirmed in figure 3. This shows the standard deviations of the normally distributed *M*-particle canonical distributions evaluated at four different values of $v_m$.

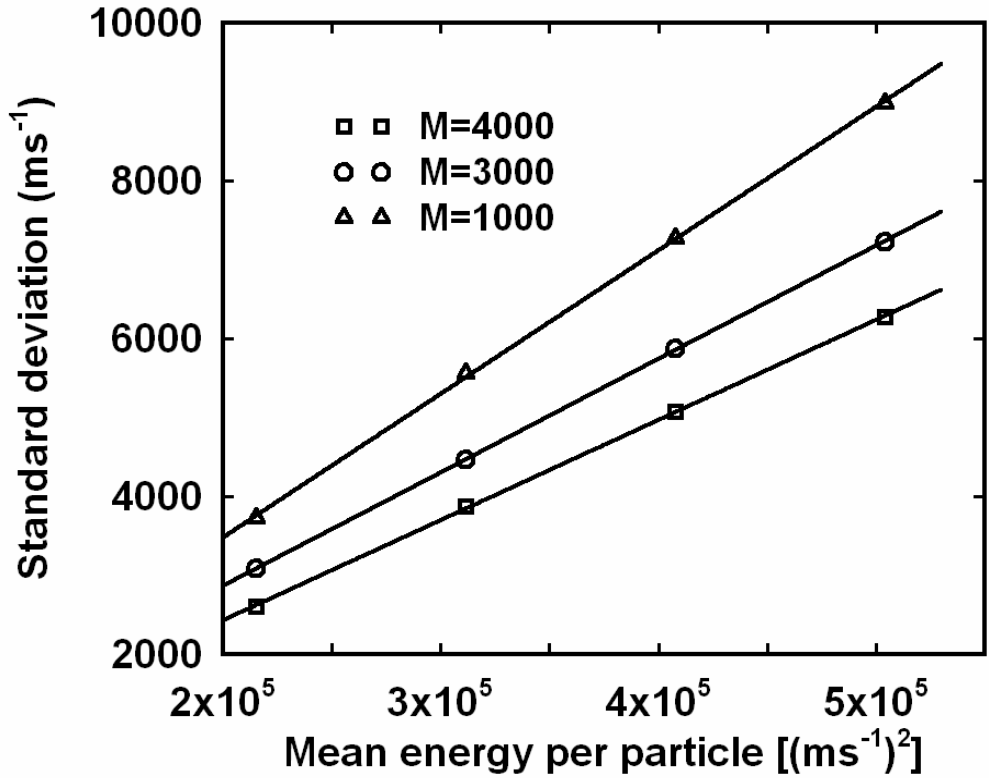

Figure 4.
The standard deviation of three normal distributions plotted against mean energy per particle as a measure of temperature.

Figure 4 shows the standard deviation against $\overline{E}$ for three different values of *M* and confirms that $\langle E_1 \rangle^2 \propto T^2$.

## Conclusion

The mathematics leading to equation (9) is all theory but does serve to illustrate the use of methods of mathematical statistics in statistical mechanics. Here the central limit theorem is used to derive a little appreciated, but very important result; the canonical distribution is not in fact the exponential distribution so often given in undergraduate texts but is a normal distribution centred on the mean energy in the system. This has been demonstrated by means of a very simple numerical simulation which, moreover, also reproduces the essential features of fluctuations in statistical thermodynamics. This shows very clearly that the canonical distribution represents in reality the tendency of the energy in the system to fluctuate about its mean value. The idea of fluctuations is also used in modern derivations of the incorrect exponential form of the canonical distribution, but the present numerical simulation clarifies the physics.

It is well known, of course, that numerical simulations can lead to important physical insights, but the present work shows how important a correct theoretical understanding is for their correct interpretation. Binder [9], for example, mentions in his extensive review of Monte Carlo techniques, that what he calls "simple sampling", which is essentially the unrestricted sampling technique employed here, can lead to normal distributions in Monte Carlo simulations of, for example, spin systems. As such distributions are at odds with the exponential form of the canonical distribution in (10), simple sampling has been rejected and replaced by "importance sampling", in which certain samples are rejected as being unimportant. If the rejection criteria are chosen correctly, distributions of energy which obey (10) rather than the normal distribution of (9) can be produced. In light of the present work, the flaws in this procedure are all too apparent, but the dilemma - should the results of numerical simulations be taken at face value or should well-established, but conflicting, theory take precedence? - is very real.

Whether "importance sampling" is valid is too general a question for this paper, but it is worth noting that the sample rejection described by Binder is based on the Metropolis criterion. This uses the Boltzmann factor to determine the probability of acceptance so perhaps it is not surprising that the distribution of accepted energy states is essentially a Boltzmann distribution. The "simple sampling" used in this work which leads to a normal distribution of energy states is undoubtedly valid. The only assumption in the numerical simulation is that the speeds should be drawn from a Maxwell-Boltzmann distribution and this is achieved through appropriate scaling rather than sample rejection. The majority of particles are found to have the most probable speed of the Maxwell-Boltzmann distribution and by definition the mean square speed corresponds to the mean energy. Low energy states, on the other hand, can only be achieved if the majority of particles have a low speed. The lowest possible energy corresponds to all the particles having zero speed, which is highly improbable. Theory requires this and it emerges naturally from the simulation. Therefore equation (10) cannot describe the canonical distribution of a classical gas and the only possible conclusion is that equation (9) gives the correct distribution. This has been derived from statistical principles rather than from consideration of the physics and thus emphasizes the importance of a sound understanding of mathematical principles.  Possibly, science students might be encouraged not to shy away from mathematical texts too much in the future.

# References.

**Appendix.**

The aim of this work was to generate the Maxwell-Boltzmann speed distribution rather than normally distributed random numbers, though of course one implies the other. An interesting observation made in the course of this work showed that if the orthogonal velocity components of the speed are distributed uniformly, the distribution of the speeds is still a peaked function (figure A1) with the distribution showing a clear parabolic dependence on speed for low values of *v*. This demonstrates in an unexpected, but very convincing manner, how the density of states depends on *v*.

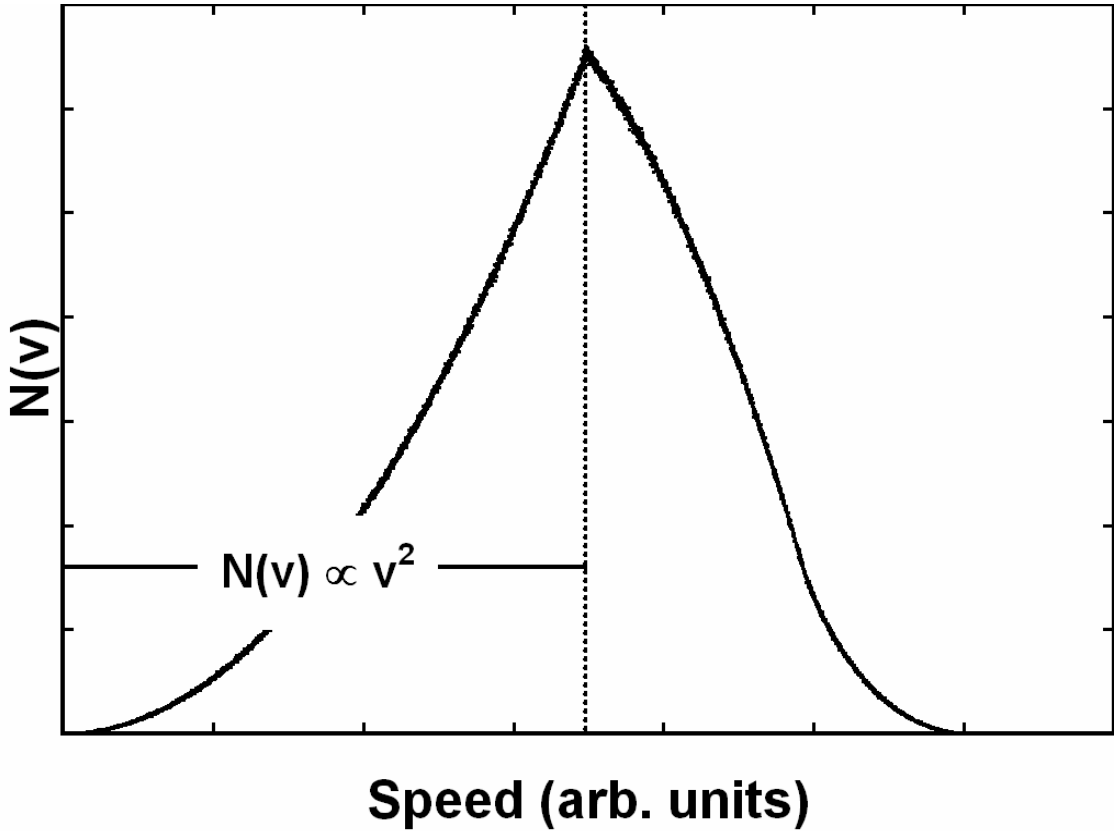

Figure A1.
The speed distribution generated by uniform scaling of randomly generated orthogonal velocity components. The parabolic dependence of the density of states is clearly demonstrated.

It is also noticeable in figure A1 that there is a large tail at high *v* despite the uniform distribution of the orthogonal components. There is also a clearly defined upper limit to the maximum speed which arises from the definite scaling limit to the random number. This suggests that if the scaling itself can be randomised the distribution will have a much smoother tail. Trial and error revealed that if

$$L = \ln[R_1] \qquad (A1)$$

where $0+\gamma \leq R_1 \leq 1$ is a uniformly generated random number and

$$f(L) = |L|^\beta \tag{A2}$$

hen a second uniformly generated random number, $R_2$, scaled between the limits $\pm v_{\text{lim}}.f(L)$ yields a number $z_j$, for which the vector sum

$$v = \left\{ \sum_{j=1}^{3} z_j^2 \right\}^{1/2} \tag{A3}$$

is distributed according to

$$N(v) = a.v^2 \exp\left[\frac{-v^\alpha}{b}\right] \tag{A4}$$

The Maxwell-Boltzmann distribution is generated for $\alpha=2$, for which $\beta=0.394476$ (figure A2).

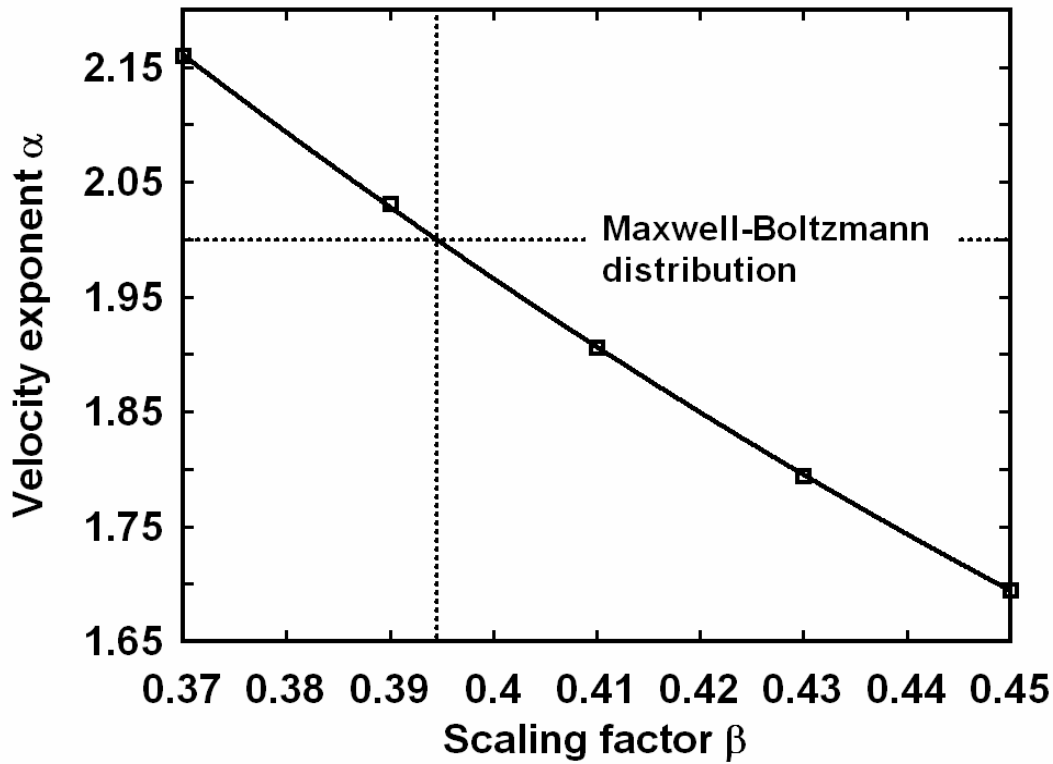

Figure A2.
The Maxwell-Boltzmann distribution is generated for a velocity exponent of $\alpha=2$ corresponding to a scaling factor of $\beta=0.394$ in equation A4.

It follows, then, from the theory of the Maxwell-Boltzmann speed distribution that $z$ is distributed normally, as illustrated in figure A3 for $v_{\text{lim}}=1$.

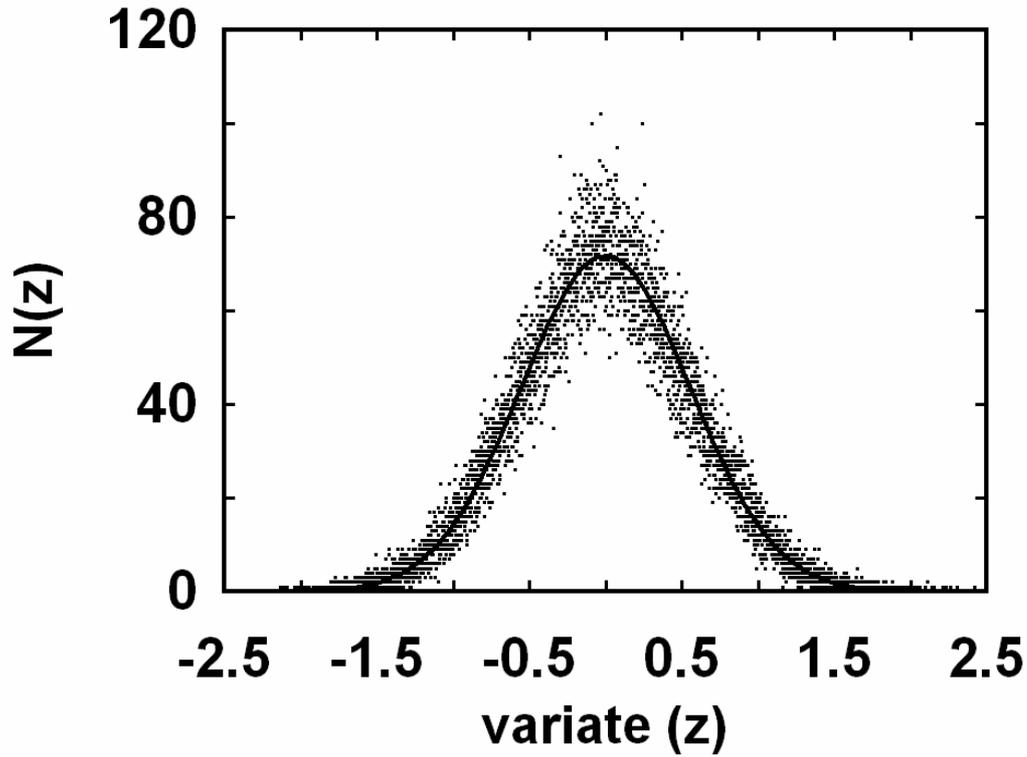

Figure A3.
The distribution of $10^4$ random numbers scaled according to the algorithm (A1)-(A4). The solid line is a best fit normal distribution.

The standard deviation is 0.55565 which allows the scaling limits to be defined in terms of the most probable velocity, $v_m$, as

$$v_{\lim} = \frac{1}{0.55565} \frac{v_m}{\sqrt{2}} \qquad (A5)$$

The parameter $\gamma$ in (A1) is necessary to prevent the occurrence of ln(0) and was chosen to be $10^{-6}$, which allows for a maximum value of $|L|=13.8$. Thus large values of $z$ can be generated but with a low probability.

This algorithm works even if the two random numbers, $R_1$ and $R_2$, are generated from the same random number generator, but we chose to use two different generating algorithms each seeded by the output of the other.